\documentclass[preprint,journal]{vgtc}            % preprint (jourto style)

\newcommand{\productname}{HaptoFloater}
\newcommand{\hapticreactor}{HAPTIC\texttrademark~Reactor\@\xspace}

\newcommand{\commentout}[1]{}

\usepackage[whole]{bxcjkjatype}

\usepackage{color}
\usepackage{savesym} \savesymbol{spacing} % to avoid conflict with setspace
\usepackage{setspace}
\definecolor{Orange}{rgb}{1,0.5,0}

\definecolor{DarkGreen}{rgb}{0,0.5,0}
\definecolor{Purple}{rgb}{0.7,0,0.7}
\definecolor{Blue}{rgb}{0.2,0.2,0.8}
\definecolor{Red}{rgb}{1.0,0.0,0.0}
\definecolor{Brown}{rgb}{0.7,0.4,0.1}

\usepackage{multicol}

\usepackage{xspace}
\newcommand{\etal}{et al.\@\xspace} % Prints ``et al.'' with proper spacing
\newcommand{\Fig}{Fig.\@\xspace} % Prints ``Fig.'' with proper spacing
\newcommand{\ie}{i.e.\@\xspace} % Prints ``i.e.'' with proper spacing
\onlineid{1080}

\vgtccategory{Research}

\title{\productname: Visuo-Haptic Augmented Reality by\\Embedding Imperceptible Color Vibration Signals for\\Tactile Display Control in a Mid-Air Image}

\author{%
  \authororcid{Rina Nagano}{0009-0007-8338-8563}, Student Member, IEEE,
  \authororcid{Takahiro Kinoshita}{0009-0000-9832-0296},
  \authororcid{Shingo Hattori}{0000-0003-4573-799X},
  \authororcid{Yuichi Hiroi}{0000-0001-8567-6947}, Member, IEEE,\\
  \authororcid{Yuta Itoh}{0000-0002-5901-797X}, Senior Member, IEEE,
  and \authororcid{Takefumi Hiraki}{0000-0002-5767-3607}, Member, IEEE
}

\authorfooter{
  \item Rina Nagano, Takahiro Kinoshita, Shingo Hattori, and Takefumi Hiraki are with University of Tsukuba. E-mail: nagano@pml.slis.tsukuba.ac.jp, kinoshita@hpcs.cs.tsukuba.ac.jp.
  \item Takahiro Kinoshita, Shingo Hattori, Yuichi Hiroi, and Takefumi Hiraki are with Cluster Metaverse Lab. E-mail: \{s.hattori, y.hiroi, t.hiraki\}@cluster.mu.
  \item Yuta Itoh is with the University of Tokyo. E-mail: yuta.itoh@iii.u-tokyo.ac.jp.
}

\abstract{
  We propose \productname, a low-latency mid-air visuo-haptic augmented reality (VHAR) system that utilizes imperceptible color vibrations. When adding tactile stimuli to the visual information of a mid-air image, the user should not perceive the latency between the tactile and visual information. However, conventional tactile presentation methods for mid-air images, based on camera-detected fingertip positioning, introduce latency due to image processing and communication.
  To mitigate this latency, we use a color vibration technique; humans cannot perceive the vibration when the display alternates between two different color stimuli at a frequency of 25 Hz or higher. In our system, we embed this imperceptible color vibration into the mid-air image formed by a micromirror array plate, and a photodiode on the fingertip device directly detects this color vibration to provide tactile stimulation.
  Thus, our system allows for the tactile perception of multiple patterns on a mid-air image in 59.5 ms.
  In addition, we evaluate the visual-haptic delay tolerance on a mid-air display using our VHAR system and a tactile actuator with a single pattern and faster response time.
  The results of our user study indicate a visual-haptic delay tolerance of 110.6 ms, which is considerably larger than the latency associated with systems using multiple tactile patterns.
}


\teaser{
  \centering
  \includegraphics[width=\linewidth, alt={teaser_image}]{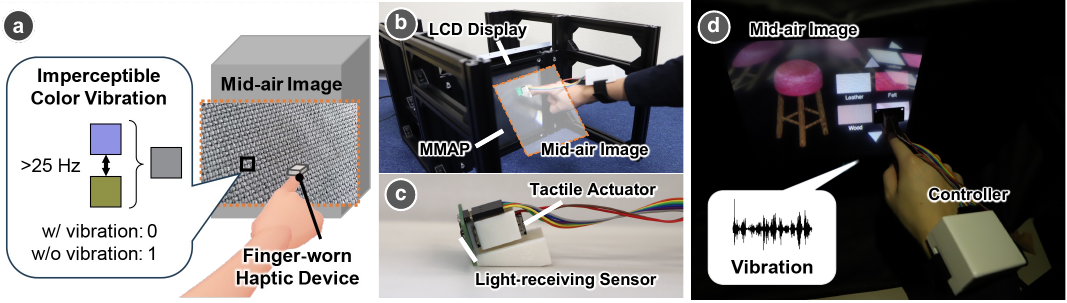}
  \caption{(a) Conceptual image of \productname. Haptic vibration control information is embedded in each pixel of the mid-air image using imperceptible color vibration. This light information controls a finger-worn haptic device. (b) Overview of the assembly of a mid-air display. It consists of a high-brightness LCD display and a micro-mirror array plate (MMAP). (c) Overview of the finger-worn haptic device. It has a light sensor for receiving color vibration at the tip and a tactile actuator to present vibration on the side of the fingernail. (d) \productname~displays a visuo-haptic texture image as an application scenario of texture design. The user can not only see the texture of a mid-air image but can also touch it.
  }
  \label{fig:teaser}
}

\graphicspath{{figs/}{figures/}{pictures/}{images/}{./}} % where to search for the images

\usepackage{tabu}                      % only used for the table example
\usepackage{booktabs}                  % only used for the table example
\usepackage{lipsum}                    % used to generate placeholder text
\usepackage{mwe}                       % used to generate placeholder figures

\usepackage{mathptmx}                  % use matching math font

\usepackage{amsmath}
\usepackage{amsfonts}
\usepackage{mathtools}

\begin{document}

%%%%%%%%%%%%%%%%%%%%%%%%%%%%%%%%%%%%%%%%%%%%%%%%%%%%%%%%%%%%%%%%
%%%%%%%%%%%%%%%%%%%%%% START OF THE PAPER %%%%%%%%%%%%%%%%%%%%%%
%%%%%%%%%%%%%%%%%%%%%%%%%%%%%%%%%%%%%%%%%%%%%%%%%%%%%%%%%%%%%%%%

\firstsection{Introduction}

\maketitle
% Tactile perception is an important factor in our perception of the material properties of physical surfaces.
% Haptic allows us to distinguish texture (roughness, hardness, etc.) by haptically exploring surfaces, which is an important component of object reality.
% This tactile presentation is essential to enhance the realism and immersion of the user experience in augmented reality (AR)~\cite{eck2015}.
% Many studies have attempted to use Visuo-Haptic AR (VHAR), where haptic information is presented over visual information.
% As for texture presentation systems in VHAR, grounded haptic devices work in conjunction with visual displays to enable haptic presentation on virtual images.

% The main limitations of these VHAR systems were that the use of grounded haptic devices limited the user's workspace to a small manipulable area and required a calibration process to align the haptic presentation with respect to the visual presentation~\cite{inami2000visuo, Sandor2007, sandor2007exploring, Wang2011, Cosco2013}.
Tactile sensation plays a crucial role in how we perceive and interact with the world around us, providing essential information about objects' texture and material properties.
In augmented reality (AR), the integration of tactile feedback with visual stimuli to create a Visuo-Haptic AR (VHAR) experience becomes paramount to achieve a high level of realism and immersion~\cite{eck2015}. This synthesis aims to allow users to see and feel virtual objects as if they were physically present, enhancing the perception of object reality through tactile exploration.

To accomplish this, VHAR systems typically consist of two main components: an optical display system to present AR visuals and a physical haptic feedback system for tactile presentation. These systems must work seamlessly to merge haptic information with visual content, providing a coherent and immersive user experience. The challenge lies in ensuring that these tactile and visual cues are perfectly synchronized, enhancing the user's interaction with virtual environments.

However, current implementations of VHAR face significant limitations. The reliance on grounded haptic devices restricts users' movements, confining them to a small area where they can interact with the virtual environment. Moreover, these systems require meticulous calibration to ensure that tactile sensations align accurately with visual presentations~\cite{inami2000visuo, Sandor2007, sandor2007exploring, Wang2011, Cosco2013}.
This alignment is critical in maintaining the illusion of reality in VHAR, but it poses practical challenges in terms of the complexity of the setup and user experience.

A promising approach to address these challenges is to use haptic displays that are controlled by light from projection-based visual displays~\cite{Miyatake}.
These VHAR systems expand the user's workspace following projection volumes, enabling tactile interactions across a broader area. Notably, they eliminate the need for external tracking systems, enhancing the ease of use and achieving high spatio-temporal consistency between visual and haptic information.

Despite these advancements, projection-based VHAR systems still face their own set of challenges. One significant limitation is that tactile interactions are confined to the real surfaces on which the images are projected. This constraint means that while the interaction area is expanded, it remains bound to the projection surface, limiting the versatility and realism of the haptic feedback in more complex virtual environments.

However, mid-air imaging technology, which allows for the display of full-color images in real space without the limitations of physical surfaces, is garnering interest. This technology enables the overlay of computer graphics on the real-world view without necessitating devices like head-mounted displays (HMDs) to be worn around the eyes.
The capability of mid-air imaging to enhance the visualization and entertainment of real-world information makes AR systems that utilize it a key player in these fields~\cite{Chan2008-zx, Ueda2015-gh}.

An inherent challenge with mid-air image displays is the lack of tactile feedback, as the images are intangible. To address this, some researchers have explored methods to integrate tactile feedback with mid-air images through additional devices. However, visual and haptic representations in these systems are often managed through uncoordinated processes that rely on external tracking systems, such as infrared sensors, to function. This reliance frequently leads to delays and a mismatch in the positioning between visual and haptic information due to the processing and synchronization requirements of the sensing technologies. Consequently, there has yet to be a VHAR system capable of delivering diverse texture presentations that are spatio-temporally consistent and unrestricted by physical display surfaces.

This paper presents HaptoFloater, a Visuo-Haptic Augmented Reality (VHAR) system that integrates tactile feedback directly into the light of a mid-air image display (\Fig\ref{fig:teaser}a).
HaptoFloater can render visual and haptic information both independently and simultaneously while ensuring high spatio-temporal consistency between these two modalities. The system employs imperceptible color vibrations to control wearable haptic devices, embedding undetectable control signals to users within the projected image. This technique allows for embedding control information on a pixel-by-pixel basis without altering the image's visual quality, ensuring that the displayed images remain comfortable for human viewing.

The haptic displays are equipped with photo sensors that detect the control signals embedded in the image's light, enabling the presentation of vibrotactile stimuli corresponding to each color channel's vibration patterns. This direct transmission of control signals through the image light allows the haptic feedback to be synchronized with the visual display with virtually no perceptible delay. Through this approach, HaptoFloater solves the previously mentioned challenges, seamlessly integrating tactile and visual information in mid-air displays.

We implemented a proof-of-concept prototype of our HaptoFloater system consisting of a mid-air image display system and a finger-worn haptic device as shown in \Fig\ref{fig:teaser}b.
% The system can embed information in eight different symbols by oscillating colors for each of the R, G, and B color channels, but due to limitations caused by human perceptual characteristics, it is known that the symbols can be used are reduced~\cite{Hattori}.
We use four vibration symbols that are suitable to use for visual applications to embed haptic control information, which is embedded into the mid-air image generated by a micro-mirror array plate (MMAP).
Within the finger-worn device (\Fig\ref{fig:teaser}c), a photodiode reads this color-vibration information and a tactile actuator presents the corresponding tactile information.
With this configuration, our system provides a correspondence between the optical information embedded in the video and the position of the fingertip, and can present tactile information with low latency and positional consistency.
% We use a tactile actuator with a wide presentable frequency range for finger-worn devices to present complex tactile textures (\Fig\ref{fig:teaser}c).

Through delay evaluation experiments in our prototype, we confirmed that our system can present visuo-haptic information with a delay of less than 100 ms, which is the reported tolerance of delay perception from vision to haptics~\cite{Miyasato1995, Silva2013-mt, Miyatake}.
This indicates that the HaptoFloater can also be used as an experimental platform to investigate human perceptual characteristics of visual-haptic perception of mid-air images.
% HaptoFloater can also be used as an experimental platform to investigate human perceptual characteristics of visual-haptic perception for mid-air images.
We evaluated the processing time (latency) of the finger-worn device from the time it receives the light of a displayed mid-air image to the tactile presentation.
In addition, we conducted user studies to determine whether the visual-haptic asynchrony (latency) of the system is acceptable to users.
Finally, several applications in daily scenes were presented (\Fig\ref{fig:teaser}d).

The contributions of this paper are that we
\begin{itemize}
\item developed a system that can superimpose tactile presentations on mid-air images by embedding imperceptible color vibration signals in the images and detecting these signals with a fingertip device.
\item conducted the first user study of visual-haptic latency perception characteristics in mid-air displays using the developed system.
\item developed three VHAR applications, an interactive touch panel, a texture design support system, and an interactive digital museum.
\end{itemize}

\begin{figure*}[t]
    \centering
    \includegraphics[width=\textwidth]{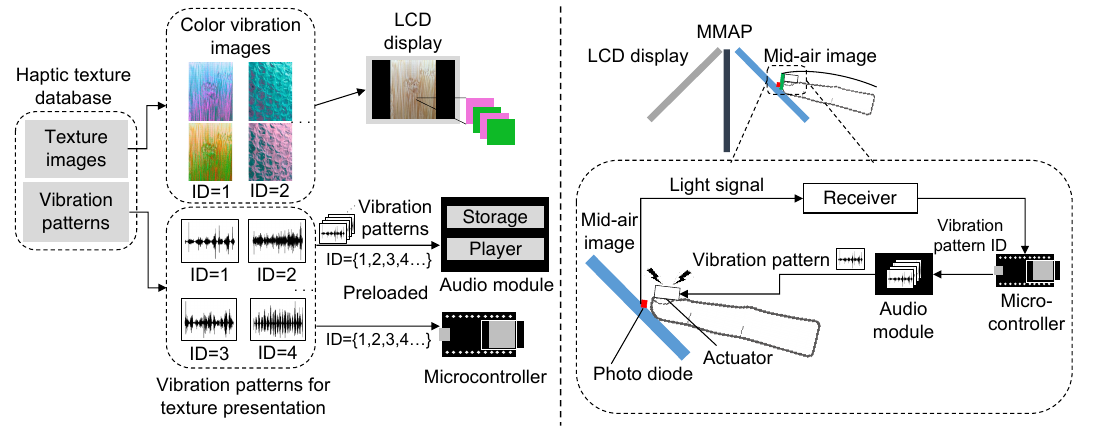}
    \caption{System principle of \productname. (Left) Processing flow of visual and tactile information. (Right) Schematic illustration of the tactile feedback mechanism.}
    % details
    \label{fig:system-overview}
\end{figure*}

% The template does not contain the respective dates of the conference/journal issue, these will be entered by IEEE as part of the publication production process.
% Therefore, \textbf{please leave the copyright statement at the bottom-left of this first page untouched}.

\section{Related Work}
This section describes previous VHAR systems that have been specifically designed for visuo-haptic texture presentation.
First, we provide an overview of existing haptic devices in VHAR and their presented limitations (Sec.~\ref{sec:related:haptic-device}).
We then explain the principles of imperceptible color vibration, the method we use to embed signals for the control of the haptic device (Sec.~\ref{sec:related:colorvib}). We also review the interface system that combines mid-air image displays and haptic devices, and discuss its potential application as a VHAR system (Sec.~\ref{sec:related:midair}). Finally, we discuss guidelines for latency from haptic to visuo-haptic information (Sec.~\ref{sec:related:perception}).

\subsection{Haptic devices in VHAR}\label{sec:related:haptic-device}
An early approach to haptic presentation was the use of grounded haptic displays such as PHANToM~\cite{phamtom}.
To use these for VHAR applications, they are usually combined with a visual display such as a half-mirror~\cite{Wang2008,inami2000visuo} or a head-mounted display~\cite{Harders2009,Sandor2007}.
These systems allow tactile information to be presented on a virtual image.
The main limitations of these systems were that the use of grounded tactile devices limited the user's workspace to a small area that could be manipulated and required a cumbersome calibration process to align the tactile presentation with respect to the visual presentation.

One way to overcome these limitations is to integrate a haptic display into a flat panel visuo-haptic display~\cite{basdogan2020, costes2020towards}.
This system ensures the spatio-temporal consistency of the visuo-haptic display by integrating touch-sensing capabilities into the display.
However, even in such visuo-haptic touch panels, the workspace is limited to a physical plane.
In contrast, wearable haptic displays, such as finger-worn devices~\cite{pacchierotti2017wearable}, styluses~\cite{lee2004haptic, kyung2008ubi}, and armbands~\cite{rahal2009continuous, he2015pneuhaptic}, enable tactile presentation without restricting the user to a small workspace.
These wearable displays can be used for both virtual object interaction~\cite{maisto2017evaluation, lee2021wearable} and tangible interface interaction~\cite{de2018enhancing, bau2012}, but require external tracking systems and tend to introduce perceptible spatiotemporal errors.

Another approach to overcome the above limitations is to combine a projection-based visual display with a wearable haptic display controlled by projected light.
HALUX~\cite{Uematsu2016-tw} is a wearable haptic display that uses the luminance of projected light to control a vibrating actuator.
In this display, the tactile actuators are controlled to move when the projected light from the projector hits them and to stop when it does not.
This display uses the projected light to enable the presentation of tactile sensations, but does not present any visuo-haptic content to the human viewer, as the projected light is only used to control the tactile display.
SenseableRays~\cite{Rekimoto2009-il} use structured light to control the vibration of a vibrating actuator.
The signal of visuo-haptic information is converted into a signal of tactile information, thereby presenting vibrotactile sensations.
The amplified received light is used as a driving signal for the actuator.
Since this system does not require an external tracking system, the tactile device can be made smaller and the latency of tactile presentation can be reduced.
The disadvantage of this system is that the visuo-haptic and haptic information cannot be designed independently.
In other words, the visuo-haptic information automatically determined the visual information because of the direct conversion between the projected light and the vibration pattern.
HaptoMapping~\cite{Miyatake} is a VHAR system that controls haptic devices using pixel-level visible light communication (PVLC), which embeds imperceptible information in the projected image.
PVLC is a technology that can embed luminance modulation information into images in a way that is invisible to humans by using a high-speed projector~\cite{Kimura2008}.
By using PVLC, this system enables independent visual and haptic design and consistent simultaneous visual and haptic presentation.
However, this system had the disadvantage that it could only work when projected images designed to embed information were projected using a special high-speed projector.

Our VHAR system uses the imperceptible color vibration~\cite{Abe2017} to embed information into the displayed image.
The imperceptible color vibration can be realized on LCD displays with a typical refresh rate of 60 Hz because the flicker frequency requirements for human perception are less stringent than those for luminance modulation.
This allows independent design of visual and haptic sensations and consistent simultaneous visual and haptic presentation using a commercial LCD display.

\subsection{Color vibration for embedding information}\label{sec:related:colorvib}
Efforts to embed information in images using color information have been studied mainly in the area of screen-camera interaction, while there are applications such as digital watermarking~\cite{Zhang2015}.
THAW~\cite{Leigh2015} tracks a smartphone with a gradient-colored 2D pattern on the display that indicates the position of the smartphone when captured by the camera.
CapCam~\cite{Xiao2016} detects cameras with capacitive touchscreens and pairs the display with the camera by sending data as a flashing color pattern.
VRCodes~\cite{Woo2012} is a technique that uses a rolling shutter smartphone camera that switches between complementary colors.
While these methods are capable of unobtrusive data transmission, they can only output in grayscale, and the codes are visible, so they are embedded exclusively in the background of the content.

Imperceptible color vibration is a method that achieves the embedding of information by switching colors with constant luminance at a rate imperceptible to humans~\cite{Yamamoto2017, Abe2017}.
This method takes advantage of the human visual property that observers perceive two intermediate colors when two different colors of the same luminance are rapidly alternated~\cite{Jiang2007-ww}.
It is known that the critical color fusion frequency (CCFF), the frequency at which humans cannot perceive color vibration, is about 25 Hz, which is about half the critical fusion frequency (CFF), the frequency at which luminance flicker becomes imperceptible.
Since the refresh rate of commercially available LCD monitors and projectors is 60 Hz or higher, the images displayed on these devices can embed information through color vibration.
The system can then send control signals to the device while displaying video content to humans.
We exploit this imperceptible color vibration to implement a low-latency, low-displacement VHAR system as a mid-air display.

\subsection{Tactile feedback on mid-air image displays}\label{sec:related:midair}
Mid-air imaging has attracted attention as a technique for displaying full-color images in real space without being constrained by physical surfaces.
Several specialized optical systems have been proposed to realize such displays, including roof mirror arrays~\cite{Maeda2014-xs}, retro-reflective mid-air imaging~\cite{Tsuchiya2021-xn}, and MMAP~\cite{Maekawa2006-al}.
In particular, MMAP can display bright images and is easy to acquire and install, so MMAP-based AR systems play an important role in real-world information visualization and entertainment~\cite{Kikuchi2022-ut, Kiuchi2022-vy}.

In mid-air image display, there is usually no haptic feedback because the images are intangible.
Therefore, research has proposed the use of ultrasonic phased arrays to add haptic feedback to mid-air images~\cite{MakinoHaptoClone, MonnaiHaptoMime, Yoshida2017-no}.
This phased array can be shaped so that the ultrasonic waves converge at this point.
This would create pressure at a specific point in the air, and if a human finger were at that point, it would be possible to present a tactile stimulus without contact.

However, a drawback of ultrasonic phased arrays is that the frequency and intensity of the pressure can only be controlled to a very limited extent compared to tactile actuators.
As a result, it is difficult to produce complex tactile sensations such as texture, although there have been efforts to expand the range of tactile sensations using ultrasound~\cite{Morisaki, Freeman}.
In addition, because the visuo-haptic and haptic presentations in these systems operate in processes that are not coordinated with each other, they require external tracking systems such as infrared sensors.
Therefore, the processing and synchronization caused by this sensing often results in latency and positional mismatch between visual and haptic information.
The delay time resulting from communication between an external tracking system and a PC is generally said to be about 100 ms or more, and light control-based devices, which in principle do not incur this delay, are supposed to have an advantage for the human interface~\cite{Hiraki2018}.

We propose a VHAR system that is spatio-temporally consistent and capable of presenting a wide variety of textures by controlling a wearable haptic display by embedding control signals in the light itself, which is composed of mid-air images.

% However, the inability to control the frequency and intensity of the pressure makes it difficult to present complex tactile sensations. While studies attempt to expand the range of tactile sensations using ultrasound~\cite{Morisaki, Freeman}, reproducing finer tactile sensations such as texture remains challenging.

\subsection{Human perception characteristics for visuo-haptic latency tolerance}\label{sec:related:perception}
When designing a VHAR system, the presented visuo-haptic stimuli must be spatio-temporally consistent to enhance the reality of the content.
Therefore, we need to know the threshold of delay perception in visuo-haptic presentation.
With regard to delay perception, Miyazato~\etal~\cite{Miyasato1995} investigated the acceptable latency range from visual to haptic sensation in a visuo-haptic teleconferencing system.
A user study showed that the threshold for delay perception in their system was about 100 ms.
Silva~\etal~\cite{Silva2013-mt} also investigated the acceptable range of perceivable latency of visuo-haptic sensation in video games. They found that the stimulus threshold for latency in their system was approximately 100 ms.
Miyatake~\etal~\cite{Miyatake} also investigated the acceptable range of visuo-haptic latency in a VHAR system with a tabletop projection display using finger-worn, stylus, and arm-worn tactile devices.
The results showed that the latency stimulus thresholds were approximately 100, 159, and 500 ms for finger-worn, stylus, and arm-mounted devices, respectively.

Previous studies have focused on investigating the tolerance of visuo-haptic latency in visuo-haptic virtual reality systems or VHAR systems that involve contact with physical surfaces.
Thus, there are still no design criteria for the consistency of visuo-haptic information in mid-air VHAR systems.
In addition, the latency threshold in VHAR systems has only been studied when the haptic device is switched from off to on.
Therefore, the stimulus threshold of latency in the VHAR system in both the on and off directions has not been investigated, and it would be interesting to understand the difference between them.
In this paper, we report the results of an experiment on the temporal latency of visuo-haptic latency by superimposing visual and haptic sensations on a mid-air image display using the prototype system we developed.

\section{\productname}
Figure~\ref{fig:system-overview} shows an overview of our \productname~system. The system comprises three components: a mid-air image display (Sec.~\ref{sec:sys-midair}), color vibration images (Sec.~\ref{sec:sys-colorvib}), and a finger-worn haptic device (Sec.~\ref{sec:sys-haptic}). The following sections describe each component in detail. We then describe how to estimate the latency generated by this system (Sec.~\ref{sec:sys-latency}).

\subsection{Mid-air image display}\label{sec:sys-midair}
\begin{figure}[t]
    \centering
    \includegraphics[width=\linewidth]{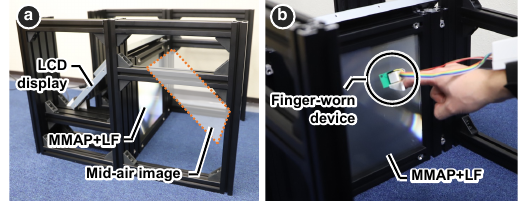}
    \caption{Mid-air display hardware of \productname. (a) Images on the LCD display screen appear in the air, symmetrical to the MMAP and the vision control film, called Louver Film (LF). (b) Touching the mid-air image with a finger-worn haptic device.}
    \label{fig:appearance}
\end{figure}

Figure~\ref{fig:appearance} shows the implementation of our mid-air image display.
We implemented a mid-air image display using an LCD (ULF1505-IX, LITEMAX, 15" diagonal, 1000 cd/m$^2$ brightness), a micro-mirror array plate (MMAP) (ASKA3D-250NT, ASKANET, 250 mm $\times$ 250 mm), and a vision control film (WINCOS Vision Control Film W-0055, LINTEC).

An MMAP behaves as a transmissive mirror that can form an image at a plane-symmetric position to the optical element.
The conceptual fabrication method for an MMAP is as follows: consider a series of mirrored glass (or acrylic) plates that are stacked on top of each other and then cut vertically into thin sheets. Given these two sheets, an MMAP can be obtained by rotating one sheet 90$^\circ$ along an axis perpendicular to its face and then gluing the two sheets together.

The vision control film, also called Louver Film (LF), is a unidirectional opaque film with opacity angles ranging from 0$^\circ$ to 55$^\circ$. It is used to shield the light emitted by the LCD and transmitted through the MMAP from the user's eyes.
In our mid-air display prototype, we placed the LCD at 45$^\circ$ to the MMAP and the LF on the opposite side of the LCD, overlapping the MMAP.

\subsection{Generating color vibration image}\label{sec:sys-colorvib}
Color vibrations alternately present two point-symmetric colors for the target color in the L*a*b* color space at constant luminance. Color pairs are quickly searched for using the method of Hattori~\etal~\cite{Hattori} to generate an array of multiple color pairs. Then, color pairs are extracted from the array that satisfies both the color vibration condition, which is imperceptible to humans, and the embedded signal condition, which allows the device to detect the vibration. These color pairs are embedded into pixels as time signals to obtain a color vibration image.

According to~\cite{Hattori}, color pairs can be generated with 9 representative colors (black, gray, white, red, green, blue, cyan, yellow, and magenta), which constitute color vibration.
Furthermore, by generating color pairs using three RGB channels, for each representative color, at least 3 types of color pairs can be obtained, and with several of representative colors, 5 types of color pairs can be obtained.
This means that, theoretically, at least 3 different vibro-tactile sensations can be presented in the representative colors, and 5 different types of vibro-tactile sensations with several of those colors.
However, in terms of implementation, we found that vibration using the G channel sometimes reproduces a grayish color. For this reason, we generated color pairs using the R and B channels, \ie, we generated 3 color pairs in which the color vibration is (on, off), (off, on), or (on, on) in the R and B channels.
These color pairs can be generated with 9 representative colors.
% In~\cite{Hattori}, these pairs satisfy the conditions for 8 representative colors excluding red.
% In~\cite{Hattori}, these pairs satisfy the conditions for 9 representative colors if we do not care about the availability of G channel vibration as in this case.

\subsection{Finger-worn haptic device}\label{sec:sys-haptic}
\begin{figure}[t]
    \centering
    \includegraphics[width=\linewidth]{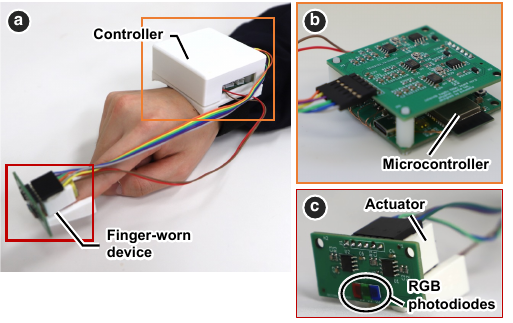}
    \caption{(a) Hardware configuration of \productname~consisting of a finger-worn haptic device and a controller unit worn on the arm. (b) The controller unit consists of an audio module that transmits tactile vibrations and a microcontroller. (c) The finger-worn haptic device consists of an actuator and light-receiving sensor board using RGB photodiodes.}
    \label{fig:hapticdevice}
\end{figure}

Figure~\ref{fig:hapticdevice} shows our implementation of a fingertip device for tactile presentation.
The sensor and actuator are controlled by a microcontroller worn on the arm (\Fig\ref{fig:hapticdevice}b).
The fingertip device (\Fig\ref{fig:hapticdevice}c) consists of a light-receiving sensor for detecting color vibrations and a tactile actuator.
The implementation details of each component are described below.

\subsubsection{Light-receiving sensor}\label{sec:impl-light-sensor}
We implemented a light-receiving sensor circuit to receive and detect color vibrations using a photodiode and an operational amplifier (op amp). 

\begin{figure}[t]
    \centering
    \includegraphics[width=0.8\columnwidth]{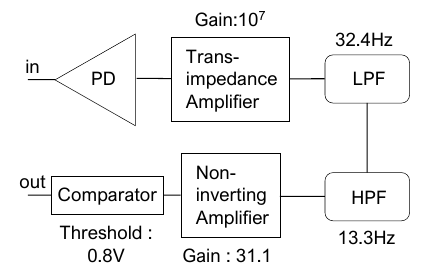}
    \caption{Configuration diagram of light-receiving sensor.}
    \label{fig:amp_f}
\end{figure}

Figure~\ref{fig:amp_f} shows a diagram of the light-receiving sensor.
The signal from each photodiode (PD) first enters a transimpedance amplifier circuit, where it is converted from current to voltage.
Then, to extract only the 30 Hz color vibration signal with high sensitivity, the signal was passed through a low-pass filter (LPF) and a high-pass filter (HPF) in sequence. 
The LPF isolates the 50/60 Hz noise from the AC power supply and the ambient light from the signal.
We implemented a 32.4 Hz Sallen-Key 4th order active LPF in the circuit.
The signal then passes through a 13.3 Hz HPF and a non-inverting op-amp to match the phase of the input and output signals.
Finally, the signal passes through the comparator to convert it to a square wave, making it easier for the microcontroller to process.

We used three photodiodes (NJL6407R, JRC) covered with an RGB color filter (NRS-1635, Nissho) as RGB photodiodes, an op-amp (AD822ARZ, Analog Devices, Inc.) for the trans-impedance amplifier, another op-amp (NJM4580E, JRC) as both the LPF and the non-inverting amplifier circuit, and a comparator (LM393DT, Texas Instruments). 
We also designed a CR-HPF circuit ourselves.

\subsubsection{Tactile actuator}\label{sec:impl-actuator}
We use \hapticreactor (Alps Alpine) for a tactile actuator embedded in the finger-worn haptic device.
\hapticreactor is a special type of linear resonant actuator (LRA), and it can convert sound transmitted through an audio amplifier into tactile vibrations, providing more complex tactile sensations over a wider frequency range than conventional LRAs.

The actuator is connected to an audio amplifier (MAX98357A, Analog Devices), and the amplifier is connected to the microcontroller (ESP32-WROOM-32E, Espressif Systems) via Inter-IC Sound (I$^2$S) interface (Fig.~\ref{fig:hapticdevice}b).

The microcontroller detects texture by the presence or absence of vibration in each of the three RGB channels.
To elaborate, each texture is represented by a unique identifier, which is associated with the pattern of presence or absence of vibration in each of the three RGB channels.
The pattern of presence or absence of vibration indicates, for instance, that there is vibration in the case of R and no vibration in the case of G and B.
Thus, by detecting the presence or absence of vibration in each of the three RGB channels, texture can be detected and identified.
However, since two channels, R and B, are used in this implementation, the texture is detected and identified by the pattern of presence or absence of vibration in each of the two RB channels.
In this way, the microcontroller inputs the corresponding ID information to the audio module based on the color vibration signals received from the light-receiving sensor circuit.
The vibration signal is stored as an audio file, and the tactile vibration presented is controlled by determining the audio file to play by the microcontroller.

For this study, texture images and the corresponding tactile vibration information (audio files) were obtained from the LMT Haptic Texture Database~\cite{StreseLMTHTB}, a tactile data set.

\subsection{Latency estimation}\label{sec:sys-latency}

%%%%%%%%%%%%%%%%%%%%%%%%%%%%%%
%%%%%  MACRO Definition  %%%%%
%%%%%%%%%%%%%%%%%%%%%%%%%%%%%%

\global\long\def\timeLate{T_{\rm total}}
\global\long\def\timeRecv{T_{\rm recv}}
\global\long\def\timeVib{T_{\rm vib}}
\global\long\def\timeThresh{T_{\rm th}}
\global\long\def\timeProc{T_{\rm proc}}

The total latency of our system $\timeLate$ is defined as the duration from the time when the haptic device is placed on a mid-air image with embedded control signals to the time when the device presents a vibration.
$\timeLate$ can be obtained by
\begin{gather}
    \label{Tlate}
    \timeLate = \timeRecv + \timeVib.
\end{gather}
The details of each duration are as follows:

$\timeRecv$ is the time from when the sensor enters the image area in which the color vibration is embedded to when the microcontroller detects the color vibration, which takes a value from 0 to half of the color vibration period, \ie, $1/(30~{\rm Hz} \times 2) = 16.7~{\rm ms}$.

$\timeVib$ is time from when the microcontroller detects the color vibration to the time the actuator begins to vibrate.
This is the sum of the processing time of the microcontroller and the audio amplifier and the mechanical response of the actuator.

\section{Evaluation}
We conducted two experiments with our system.
First, we measured the worst-case latency between visual and tactile sensations in our system (Sec.~\ref{sec:eval-system}).
Then, we evaluated the threshold at which users perceive latency between visual and tactile sensations (Sec.~\ref{sec:eval-latency}).
Finally, we evaluated the reality of the tactile sensation reproduced by our device (Sec.~\ref{sec:eval-texture}).

% The protocol of the user studies was approved by (omitted for double-blind review), and informed consent was obtained from all subjects.
The protocol was approved by the Ethical Review Committee of the University of Tsukuba Library, Information and Media Science (Registration number: 24-30) and Cluster, Inc. Research Ethics Committee (Registration number: 2023-004), and informed consent was obtained from all subjects.

\subsection{Worst-case latency of haptic device}
\label{sec:eval-system}
\begin{table}[t]
  \caption{Worst-case (Maximum) latency of the haptic from visual sensations using the finger-worn haptic device in the turn-on and turn-off conditions.}
  \label{table:maximumlatency}
  \centering
  \begin{tabular}{cccccc}
    \toprule
    & $\timeLate$ & = & $\timeRecv$ & + & $\timeVib$  \\
    \midrule
    turn-on & 59.5 ms & \ & 16.7 ms & \ & 42.8 ms \\
    \midrule
    turn-off & 46.5 ms & \ & 16.7 ms & \ & 29.8 ms \\
    \bottomrule
  \end{tabular}
\end{table}

We estimate the worst case of $\timeLate$ with our proposed finger-worn haptic device.
The measurements were taken under conditions in which the actuator was switched from off to on (called ``turn-on'' condition) and from on to off (called ``turn-off'' condition).
We measured $\timeVib$ using a \textmu s-order precision timer inside the microcontroller.
In the turn-on condition, the timer started when the microcontroller detected the color vibration and stopped when the actuator started to vibrate.
In the turn-off condition, the timer started when the microcontroller stopped detecting color vibration and stopped when the actuator ceased vibration.
The onset of vibration was detected by an acceleration sensor (KXR94-2050, Kionix) attached to the fingertip side of the device.
We performed the measurements 100 times using a microcontroller for each condition in the turn-on condition and turn-off condition.

Table~\ref{table:maximumlatency} summarizes the maximum $\timeRecv$ and $\timeVib$ for each condition, and the maximum $\timeLate$ as the sum of these latencies.
In the turn-on condition, the mean value of $\timeVib$ was 42.8 ms with a standard deviation of 0.47 ms.
In the turn-off condition, the mean value of $\timeVib$ was 29.8 ms with a standard deviation of 0.82 ms.
Since maximum $\timeRecv =$ 16.7 ms (Sec.~\ref{sec:sys-latency}), the worst case for the total system latency $\timeLate$ is 59.5 ms and 46.5 ms for the turn-on and turn-off conditions, respectively.

The method using PVLC has a delay of 96.3 ms~\cite{Miyatake}, thereby indicating that our method has a shorter delay time.

\subsection{User study on latency perception}\label{sec:eval-latency}
We conducted a user study to investigate whether users would notice a latency between visual and tactile sensations while using the system.

\subsubsection{Experimental setup}
In this experiment, we investigated the latency threshold between visual and haptic perception by introducing a delay time in the microcontroller.
The image size of the mid-air image is 150 mm $\times$ 70 mm and is lined with gray and green regions side by side.

We conducted two types of perceptual threshold assessments: perceptual threshold for delay when changing from no tactile information to providing tactile information (turn-on condition) and vice versa (turn-off condition). By incorporating a sleep time on the microcontroller in addition to $\timeLate$, the delay time could be adjusted to ensure that the system correctly provided tactile feedback after the delay time specified by the control signal ($\geq$ 59.5 ms in the turn-on condition and $\geq$ 46.5 ms in the turn-off condition).

% Under turn-on conditions, participants move their finger from the gray area to the green area. Once the finger is in the green area, the system presents the tactile sensation over a specified delay. The user reports whether or not they felt the delay. The experiments in the turn-off condition were the same except that subjects moved their fingers in the opposite direction.

% By embedding the color vibration in the blue channel of the pixel values in each area, we turned on the tactile actuator in the green area after a certain delay time and turned it off in the gray area.

\subsubsection{Participants and experimental methods}
\begin{figure}[t]
    \centering
    \includegraphics[width=\linewidth]{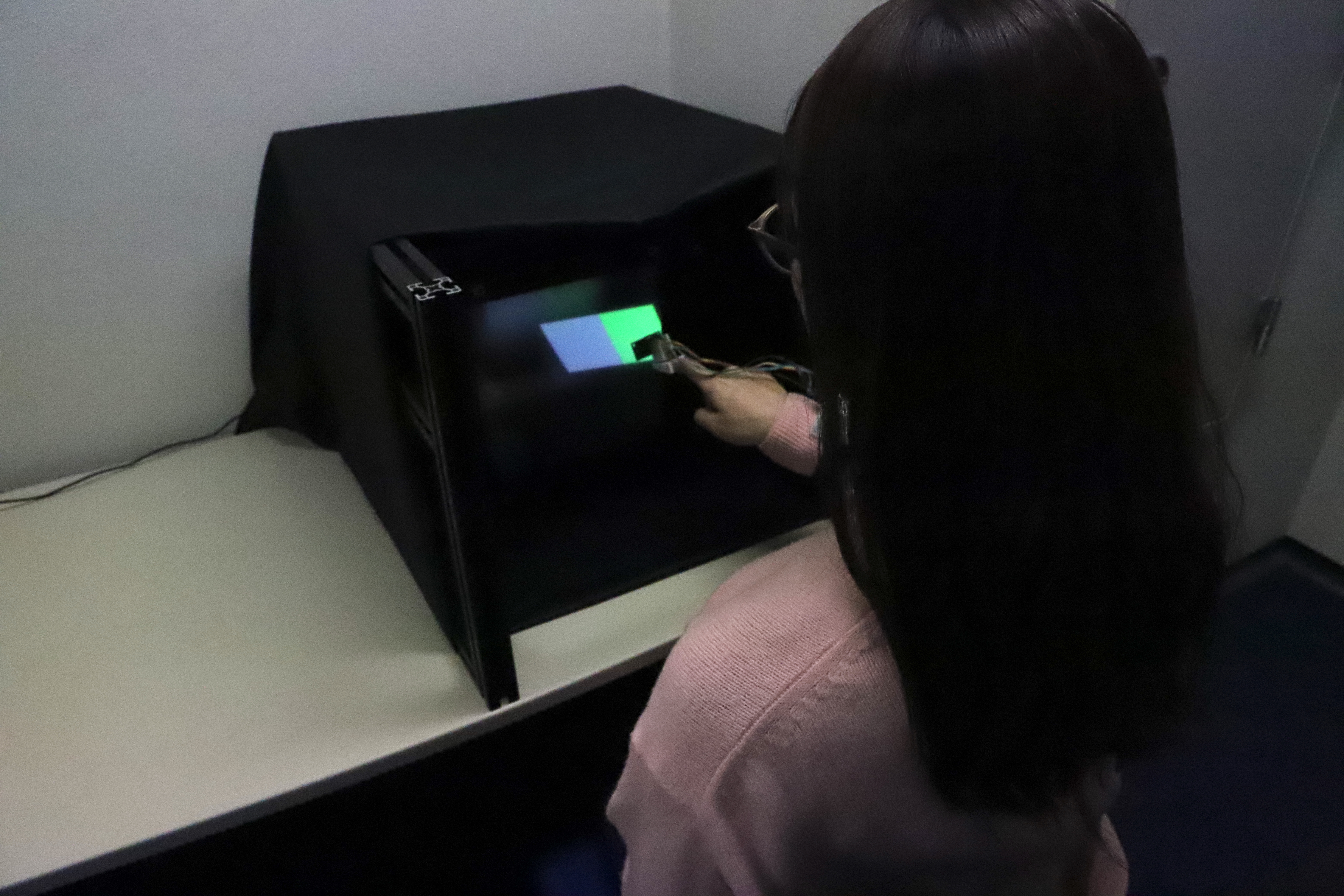}
    \caption{Experimental situation during the user study on latency perception. Although the figure shows the room with the lighting on for clarity, the actual study was conducted with dimmed lighting.}
    % The participant move their finger from the gray area to the green area and report whether they perceived a delay.}
    %Although the figure shows the room with the lighting on for clarity, the actual study was done with dimmed lighting.}
    \label{fig:setup_userstudy_latency}
\end{figure}

14 people (7 males and 7 females, aged from 20 to 24, 12 right-handed and 2 left-handed)  participated in this user study.
Participants wore the haptic device on the index finger of their dominant hand.
We prepared eight variations of latency (60, 80, 90, 100, 110, 120, 130, and 150 ms) for each of the turn-on and turn-off conditions, and these latencies also include device-dependent latencies.
Figure~\ref{fig:setup_userstudy_latency} shows the experimental situation.
During the turn-on condition, participants were instructed to move their hands from the gray area to the green area and report whether they perceived a delay corresponding to the tactile sensation from the moment they visually confirmed that their fingertips had entered the green area. 
In the turn-off condition, participants were instructed to move their hands from the green area to the gray area and report whether they perceived a delay from the moment they visually confirmed that their fingertip had entered the gray area until the tactile presentation ceased.
To control the speed of the hand movements, the experiment participants were shown the experimenter moving her hand at a speed of 150 mm/s before the commencement of the experiment. Furthermore, a metronome was played at a rate of 60 beats per minute throughout the experiment, and participants were instructed to trace the image from one end of the area to the other.
Participants rated 8 $\times$ 2 = 16 conditions per trial, with each trial repeated 8 times for a total of 128 ratings.
The order of the conditions was randomized per trial and per participant.

\subsubsection{Results}
\begin{figure}[t]
    \centering
    \includegraphics[width=\linewidth]{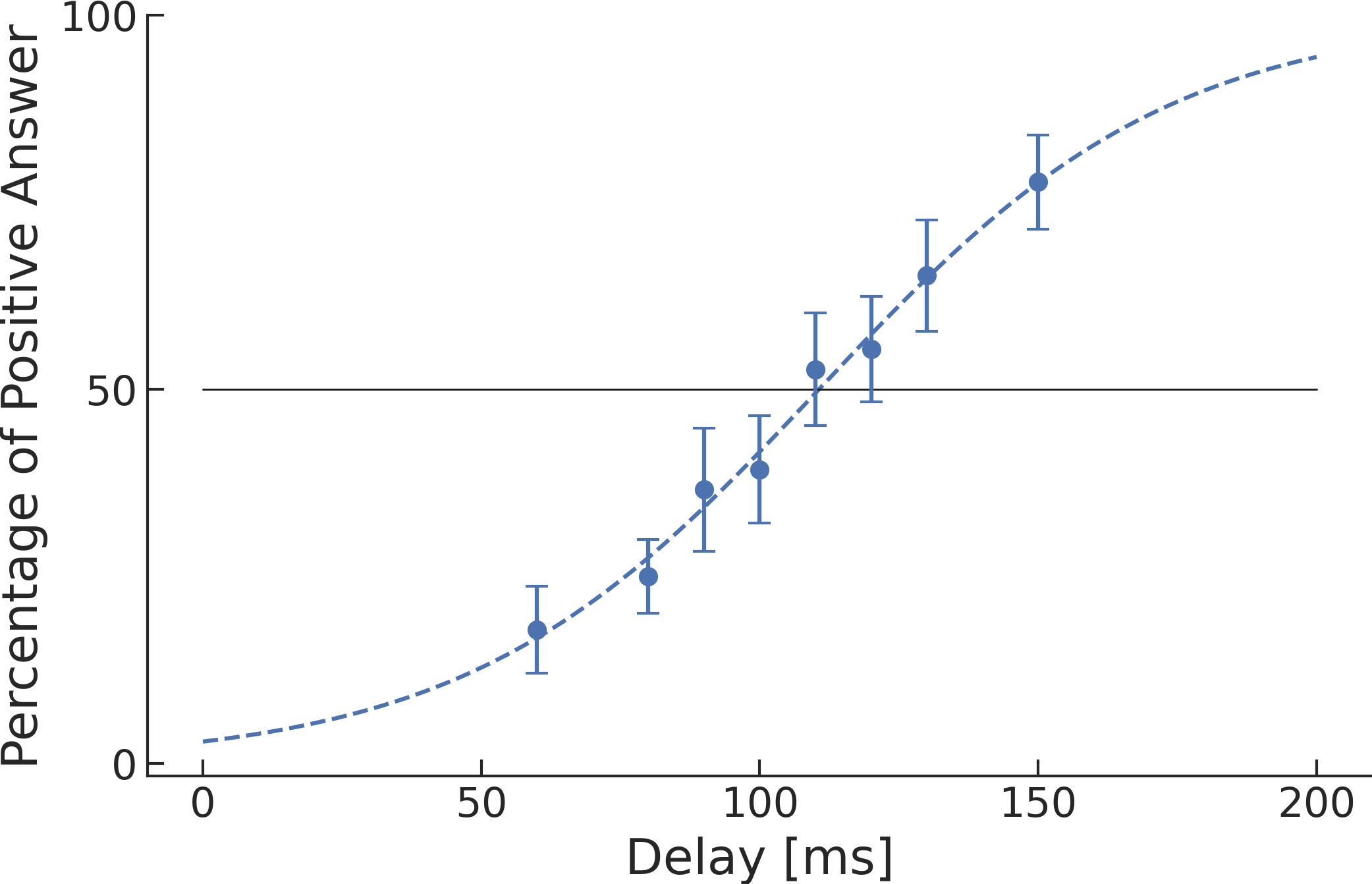}
    \caption{Percentages of positive answers in the experiment for a threshold time of a visual-haptic delay in the turn-on condition. The delay time perceived by users with 50\% probability was found to be approximately 110.6 ms.}
    \label{fig:result_userstudy_off_on}
\end{figure}

\begin{figure}[t]
    \centering
    \includegraphics[width=\linewidth]{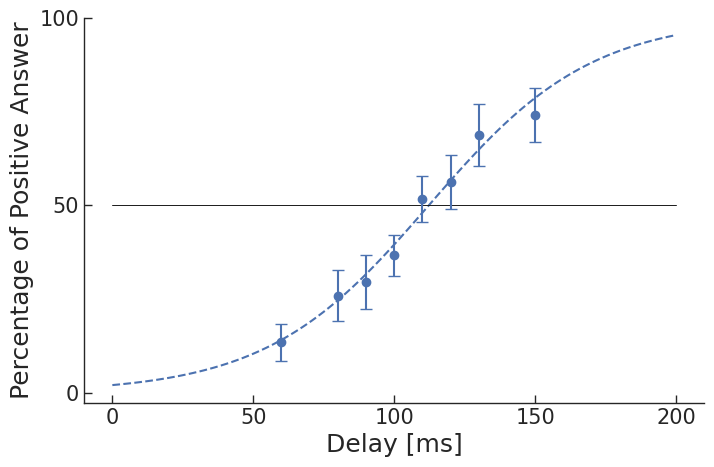}
    \caption{Percentages of positive answers in the experiment for a threshold time of a visual-haptic delay in the turn-off condition. The delay time perceived by users with 50\% probability was found to be approximately 112.3 ms.}
    \label{fig:result_userstudy_on_off}
\end{figure}

Figure~\ref{fig:result_userstudy_off_on} and Fig.~\ref{fig:result_userstudy_on_off} show the averaged percentage of positive answers obtained in the user study for each of the turn-on and turn-off conditions, respectively.
The bars represent the standard error of the mean, and the curve is fitted using the sigmoid function defined as follows:
\begin{gather}
    \label{eq:sigmoid}
    y = \frac{1}{1+\exp(-k(x-x_0))} \times 100.
\end{gather}
The fitting resulted in parameter values of $k = 0.0316$ and $x_0 = 110.64$ in the turn-on condition, and $k = 0.0346$ and $x_0 = 112.33$ in the turn-off condition, which were used in the calculations.
We identified the threshold values ($\timeThresh$) for the acceptable delay time for visual and tactile perception.
This was defined as the time at which the user perceived a delay with a 50\% probability.
The results indicated $\timeThresh$ in the turn-on and turn-off conditions are approximately 110.6 ms and 112.3 ms, respectively.
In previous studies, a visual-haptic delay tolerance for users viewing a virtual finger in a display image using a finger-worn force feedback device~\cite{Silva2013-mt} and the delay tolerance when viewing the user's own fingertip on a tabletop display using a finger-worn haptic device~\cite{Miyatake} have been reported, both of which are approximately 100 ms.
These values are about 10 ms less than the latency tolerance $\timeThresh$ obtained for the mid-air display.
This suggests that the latency tolerance is slightly looser than in the previous studies because VHAR with mid-air images involves moving the finger in the air, where there are no physical cues.

Since these latency thresholds are significantly larger than $\timeLate$, the proposed finger-mounted haptic device meets the requirement of having a latency that does not allow the user to perceive the visuo-haptic latency.
Thus, our system maintains temporal consistency between visual and haptic modalities.
This measured threshold can also be used as a design criterion for temporal consistency when considering other configurations of our system, such as the use of different actuators.

\subsection{User study on texture perception}\label{sec:eval-texture}
We conducted a user study to investigate how the user's perception of the reality of the texture varies depending on the presence and appropriateness of the tactile presentation when the proposed VHAR system is used to display visuo-haptic textures.

\subsubsection{Experimental setup}
\label{sec:eval-texture-setup}
In this experiment, we investigated how the user's perception of texture reality changes when the type of tactile sensation is changed in a visuo-haptic texture presentation.
The mid-air images were 120 mm $\times$ 70 mm in size, and five different textures (Brick, Cork, Sheep Skin, Paper, Jeans) were selected from the LMT Haptic Texture Database~\cite{StreseLMTHTB}.
In selecting the textures, we referred to the ``perceptual similarity between all our surface materials'' information in the database and select five textures from each of the nine categorized groups.
We used three types of tactile presentation: no tactile sensation (only a texture image), a single frequency (150 Hz) vibration, and a texture vibration.
For texture vibration presentation, we used audio files provided by the LMT Haptic Texture Database.

The audio files from the database were normalized so that the maximum/minimum values were $\pm$ 8192, which is one quarter of the maximum/minimum recorded value in a 16-bit wav file.
In addition, the amplitude of the single frequency audio file was also generated to be 8192.
This process was done to match the volume levels of the database audio file and the single-frequency audio file.
We did not include a sleep time in the microcontroller in this experiment, so the latency of the visuo-haptic latency is expected to be $\timeLate$.

\subsubsection{Participants and experimental methods}
\begin{figure}[t]
    \centering
    \includegraphics[width=\linewidth]{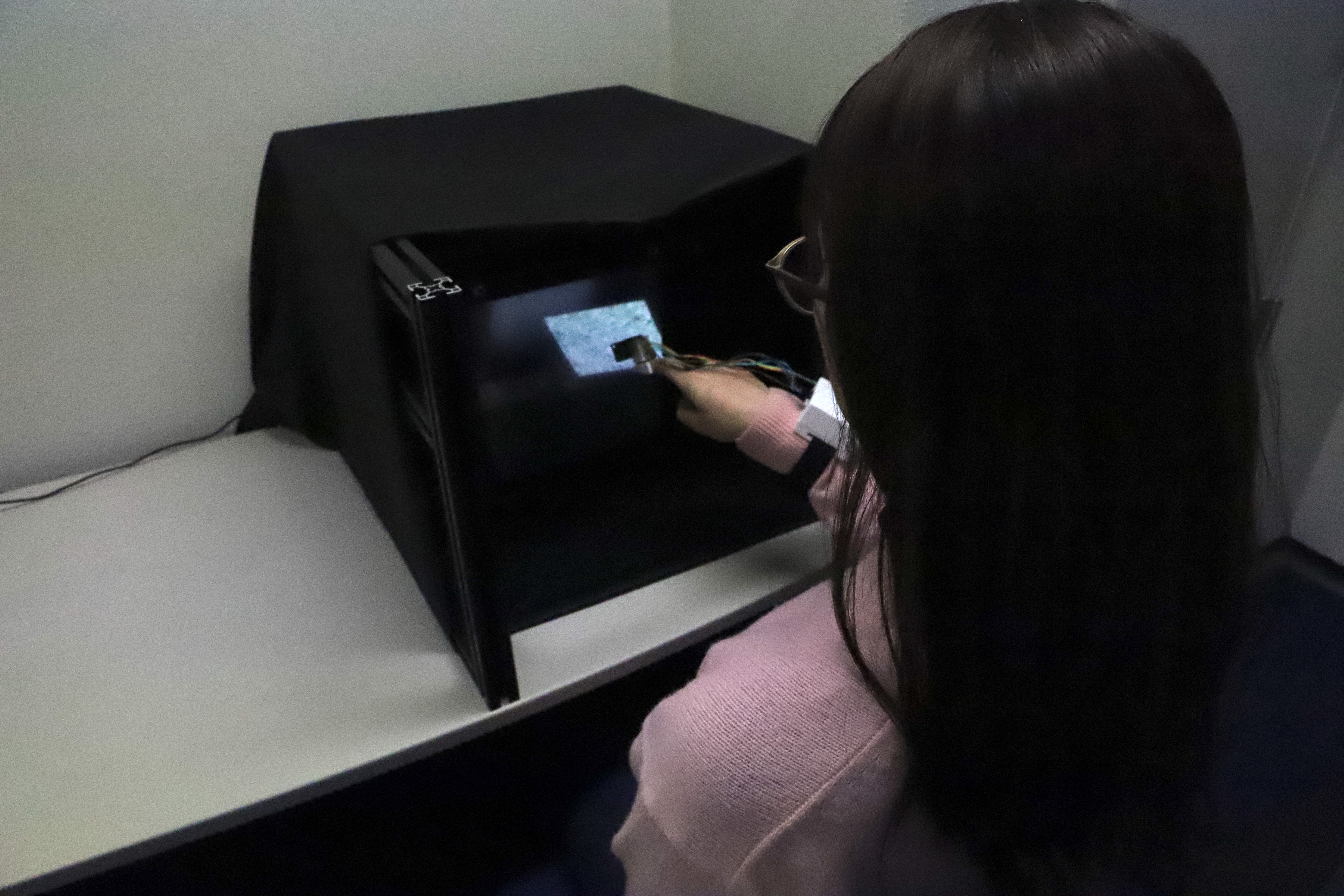}
    \caption{Experimental situation during the user study on visuo-haptic texture perception. The participant touches the mid-air image and rate how realistic the material was.}
    %Although the figure shows the room with the lighting on for clarity, the actual study was done with dimmed lighting.}
    \label{fig:setup_userstudy_realistic}
\end{figure}

\begin{figure*}[t]
    \centering
    \includegraphics[width=\textwidth]{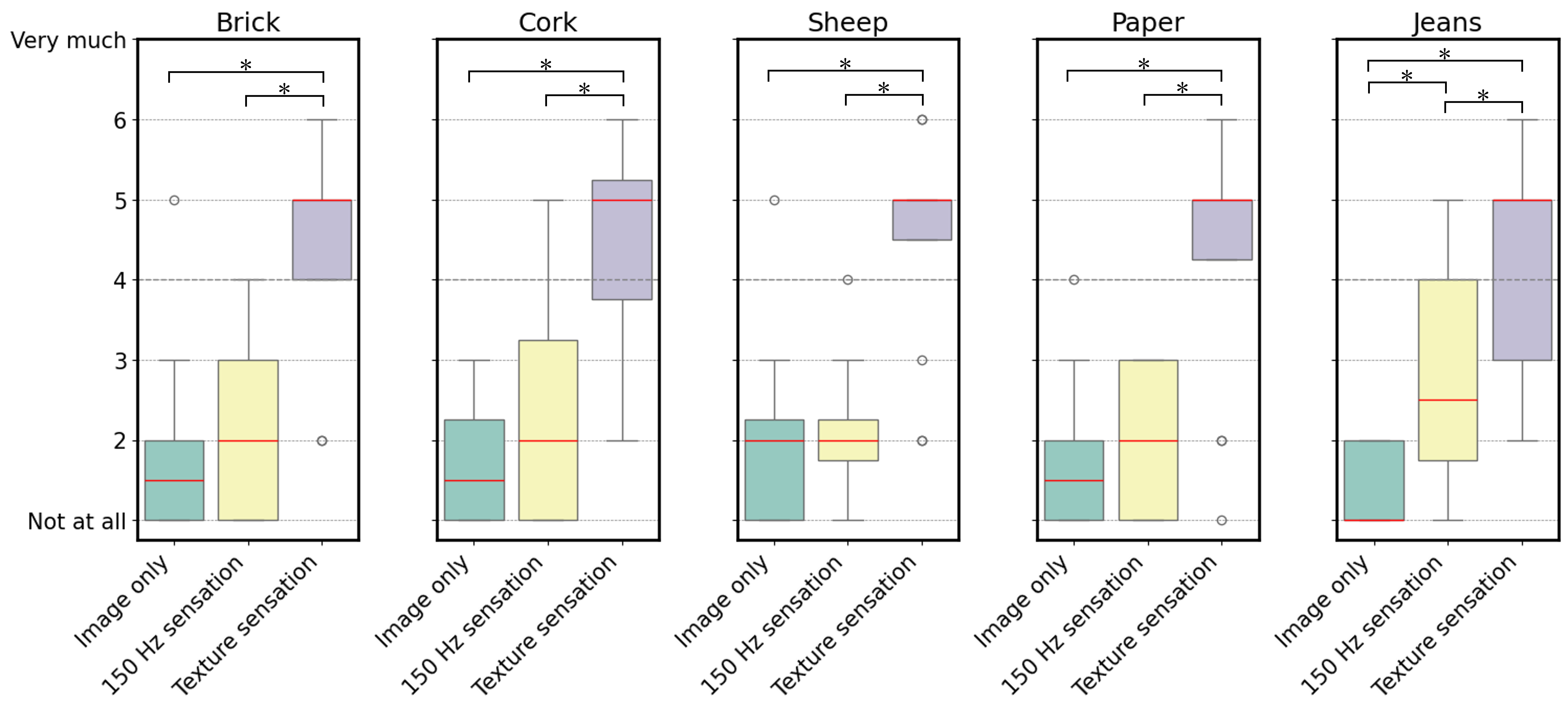}
    \caption{User ratings of texture tactile reality on a 7-point Likert scale, where image only is when no tactile sensation is presented, 150 Hz sensation is when a sine wave is presented, and texture sensation is when texture tactile sensation is presented.}
    % details
    \label{fig:result_userstudy_realistic}
\end{figure*}
Figure~\ref{fig:result_userstudy_realistic} shows the experimental result.

% \begin{table}[t]
%   \caption{Worst-case (Maximum) latency of the haptic from visual sensations using the finger-worn haptic device in the turn on and turn off conditions.}
%   \label{table:maximumlatency}
%   \centering
%   \begin{tabular}{cccccc}
%     \toprule
%     & $\timeLate$ & = & $\timeRecv$ & + & $\timeVib$  \\
%     \midrule
%     turn on & 59.5 ms & \ & 16.7 ms & \ & 49.8 ms \\
%     \midrule
%     turn off & 46.5 ms & \ & 16.7 ms & \ & 29.8 ms \\
%     \bottomrule
%   \end{tabular}
% \end{table}

12 participants (7 males and 5 females, aged 20 to 24 years, 12 right-handed) participated in this user study.
Participants wore the tactile device on the index finger of their dominant hand.
We prepared five textures, as described in Sec~\ref{sec:eval-texture-setup}.
Figure~\ref{fig:setup_userstudy_realistic} shows the experimental situation.
Participants experienced the texture presentation to their satisfaction by moving their hands in the mid-air area where the texture image was displayed.
After each experience in a single condition, participants answered the question ``How realistic was the texture material?'' on a 7-point Likert scale (1: not at all, 7: very much).
Participants evaluated 3 conditions $\times$ 5 textures = 15 conditions.
The order of texture presentation was randomized per participant and the order of presentation conditions was randomized per texture.

\subsubsection{Results}

\begin{figure*}[t]
\centering
\includegraphics[width=\linewidth]{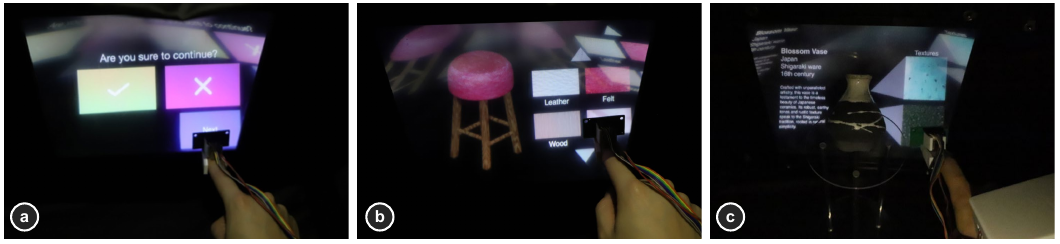}
\caption{(a) Mid-air touch panel with multiple tactile feedback. When the user touches a floating button with a finger close to it, they immediately perceive a different click feeling (tactile feedback) for each button. (b) Texture design support system. The user can see the visual texture as a mid-air image and experience the corresponding tactile feedback at precisely the appropriate timing and position. (c) Tactile presentation in museums. The vase the user is looking at is a real object, and the mid-air image is presenting VHAR from above it. The user can confirm the texture of the vase by touching a magnified mid-air image of the vase with a fingertip device.}
\label{fig:applications}
\end{figure*}

\global\long\def\pBrick{p_{\mathrm{Brick}}}
\global\long\def\pCork{p_{\mathrm{Cork}}}
\global\long\def\pSheep{p_{\mathrm{Sheep}}}
\global\long\def\pPaper{p_{\mathrm{Paper}}}
\global\long\def\pJeans{p_{\mathrm{Jeans}}}

Figure~\ref{fig:result_userstudy_realistic} shows user ratings of texture tactile reality on a 7-point Likert scale, where image only is when no tactile sensation is presented, 150 Hz sensation is when a sine wave is presented, and texture sensation is when texture tactile sensation is presented.
For Image Only, 150 Hz Sensation, and Texture Sensation, in that order, the mean and standard deviation are as follows: For brick, the mean and the standard deviation were (1.83, 1.14), (2.17, 1.14), and (4.42, 1.19), respectively. Similarly, for cork, they were (1.75, 0.83), (2.42, 1.32), and (4.50, 1.38); for sheep skin, they were (2.00, 1.15), (2.08, 0.86), and (4.50, 1.32); for paper they were (1.75, 0.92), (2.00, 0.91) and (4.33, 1.60); for jeans they were (1.42, 0.49), (2.75, 1.42) and (4.17, 1.28).

We conducted a Shapiro-Wilk test on each of the data sets to ascertain the normality of the data. The results demonstrated that none of the experimental data with the five samples exhibited normality.
Therefore, a Friedman's test, a non-parametric test, was performed. The results indicated that there were significant group differences in all experiments (p < .001).
A Wilcoxon signed-rank test (Shaffer-corrected) was then performed on all combinations of the two groups as a subtest.
The results showed that the texture vibration condition was significantly higher than the no tactile sensation condition ($\pBrick$ = .003, $\pCork$ = .002, $\pSheep$ = .009, $\pPaper$ = .003, $\pJeans$ = .006) and significantly higher in the texture vibration condition than in the single frequency vibration condition ($\pBrick$ = .006, $\pCork$ = .006, $\pSheep$ = .003, $\pPaper$ = .009, $\pJeans$ = .045). 

Furthermore, for the Jeans texture only, the condition presenting single-frequency vibration was significantly higher than the condition not presenting tactile sensation ($\pJeans$ = .011), but no significant differences were observed for the other textures.

Thus, we found that presenting texture vibration is an effective method for presenting realistic visuo-haptic textures.
\section{Applications}
We have developed three potential applications that use \productname.

\subsection{Mid-air touch panel with multiple tactile feedback}

Figure~\ref{fig:applications}a shows a user interface system that allows users to experience a variety of tactile feedback while retaining the benefits of a mid-air touch panel, providing a richer experience.
Specifically, the application allows users to feel tactile stimuli of different textures depending on where they touch when operating a button.
Because this system does not limit the number of people using the system and is not a physical touch screen, it reduces the risk of infection from sharing a touch screen with multiple unidentified users, such as in public places.
In addition, since no fingerprints are left on the screen, secure button operation can be achieved.
While retaining the benefits of the mid-air touchscreens, this system provides a richer experience with a variety of tactile feedback.

\subsection{Texture design support system}
Figure~\ref{fig:applications}b shows a texture design support system for physical surfaces.
In this application, the signal is embedded in the texture image, and the corresponding tactile pattern is preloaded into the audio module.
As the user touches and explores the surface, the tactile device selects the tactile pattern corresponding to that location and presents it to the user.
It is hoped that users will benefit from this system when searching for desired textures from a group of samples for a new product.
Since the texture is presented only as a mid-air image, the user can easily try different textures by changing the image and vibration pattern.
This system will help the user to test textures and find the right texture without scattering physical texture samples all over the workspace.

\subsection{Tactile presentation in museums}
Figure~\ref{fig:applications}c shows an example of an application for tactile presentation of exhibited objects in a museum. The mid-air image can present information associated with the real object while simultaneously viewing the real object beyond the image. By adding tactile information to the mid-air image in addition to visual information, the user can experience the tactile sensation of the texture of the exhibited object, which cannot actually be touched.

\section{Discussion}
\subsection{Perception of image flicker}
In the experiment, two-color images were displayed alternately at 60 Hz, resulting in color vibration at 30 Hz.
Color fusion occurs at this 30 Hz color vibration because the critical fusion frequency of colors in the human eye is about 25 Hz~\cite{Jiang2007CCFF}.
%Furthermore, in this paper, the color vibration images were generated using a La*b* color space designed to approximate human vision.
Furthermore, in this paper, we used a La*b* color space designed to approximate human vision, based on the previous work~\cite{Hattori}, to produce color vibration images with stabler values of L, the luminance component.
Therefore, it is expected that humans will not perceive the flicker.

However, in our experiments, flickering of the color vibration image was sometimes observed.
The experimental results of the study~\cite{hattori2024} indicate that approximately 10\% of users may experience flicker caused by the color pairs utilized in the user experiments on latency perception.
Although the luminance of the two color vibration images was adjusted equally, the perception of flickering is believed to be caused by differences in the critical fusion frequency of the colors depending on the environment of the light source, the type of color, and the individual.
Therefore, subject experiments should be conducted for each type of color.

On the other hand, many LCDs with refresh rates higher than 60 Hz, such as 120 Hz and 144 Hz, are now commercially available, and mLED displays that achieve higher refresh rates by representing gradation through high-speed LED blinking are also becoming more popular.
Using these devices, it seems possible to solve this issue by embedding information in images by means of color vibration faster than 30 Hz.

\subsection{Spatial extent of user location detection}
The precise moment at which a specific mid-air image is deemed to have been touched is when the fingertip sensor receives light from the mid-air image.
For example, in the case of the mid-air touch panel with multiple tactile feedback in Applications, the button is considered to have been touched when the fingertip sensor receives light from the mid-air image.
Due to the nature of the mid-air image display, the light is received in the vicinity of the mid-air image plane.
However, from an interaction perspective, it is necessary to determine the exact moment when the button is pressed.
Therefore, it is necessary to ascertain the detection performance at different positions in space.
It is currently understood that, due to the nature of mid-air image displays, light is more readily received at the rear of the mid-air image and less so at the front, as light tends to diverge.
Nevertheless, future investigations should be conducted to ascertain the detection performance with respect to light, depth, and other relevant factors.

% \subsection{Latency of Finger-Worn Haptic Device}
% The threshold for acceptable delay time between visual and tactile modalities in the mid-air image display system ($\timeThresh$) was approximately 107.9 ms, which is similar to the delay time ($\timeLate$) of our finger-mounted tactile device (Device HR), 108.8 ms.
% This result indicates that our haptic device can be used in a VHAR system with mid-air images without any difficulties, but there is still potential for performance improvement.

% The majority of the $\timeLate$ is accounted for by the control time of the external audio playback module.
% The audio module we used incorporates an audio playback IC capable of hardware encoding of mp3/wav.
% Some audio playback ICs guarantee playback processing with a control time of about 50 ms\footnote{\url{https://cdn.hackaday.io/files/936504006721600/YX5200-24SS\%C3\%8A\%C2\%B9\%C3\%93\%C3\%83\%C3\%8B\%C2\%B5\%C3\%83\%C3\%B7\%C3\%8A\%C3\%A9V1.6.zh-CN.en.pdf}}, so replacing the module with one that uses these ICs will reduce the latency.
% Also, using a higher-performance microcontroller and realizing audio playback processing in the microcontroller can also be expected to reduce the latency.

\section{Conclusion}
In this paper, we propose \productname, a novel visual-haptic mid-air image display that can eliminate visual-haptic delay perceived by the users.
We constructed an mid-air image display by combining an LCD and a micro-mirror array plate (MMAP), and embedded tactile control information in this image using invisible color vibration.
We also implemented a finger-worn haptic device using a tactile actuator, a light-receiving sensor board, and a microcontroller, and used the color vibration information to control this device.
We evaluated the performance of the finger-worn haptic device and showed that it could present tactile vibrations to the user with a delay of 59.5 ms after the fingertip entered the region of a mid-air image.
We conducted the user study and found that the visual-haptic delay tolerance for a finger-worn haptic device in a mid-air display is 110.6 ms.
From this result, we can conclude that our proposed finger-worn haptic device has a practical performance in visual-haptic mid-air displays.
As future work, we will investigate the visual-haptic delay tolerance for other forms of haptic devices, such as stylus-type devices, as well as finger-worn device.
In addition, we can explore more application scenarios using \productname.

% \acknowledgments{ The authors wish to thank A, B, and C. This work was supported in part by a grant from XYZ.}
\acknowledgments{
This study was supported by JST FOREST Grant Number JPMJFR206E, JST ACT-X Grant Number JPMJAX190O, JST ASPIRE Grant Number JPMJAP2327, and JSPS KAKENHI Grant Number JP20H04222, Japan.
}

% \section{Hyperlinks and Cross References}

% The style uses the \verb|hyperref| package which can typeset clickable hyperlinks using \verb|\href{...}{...}|, hyperlinked URLs using \verb|\url{...}|, and turns references into internal links.

% The style also uses \verb|cleveref| to automatically and consistently format cross references.
% We recommend that you use the \verb|\cref{label}| and \verb|\Cref{label}| calls instead of \verb|Figure~\ref{label}| or similar.
% \verb|\Cref| should be used when starting a sentence to spell out the reference (e.g.\ ``Section'') while \verb|\cref| should be used when referencing within a sentence to abbreviate (e.g.\ ``Sec.'').
% Here are examples for use within a sentence: \cref{fig:vis_papers}, \cref{tab:vis_papers}, \cref{sec:supplement_inst,sec:references_inst}, \cref{eq:sum}.
% The following sentences all start with a reference, so use \verb|\Cref|.
% \Cref{fig:vis_papers} is a \verb|figure| environment.
% \Cref{tab:vis_papers} is a \verb|table| environment.
% \Cref{sec:supplement_inst,sec:references_inst} are \verb|section| environments.
% \Cref{eq:sum} is an \verb|equation| environment.

\bibliographystyle{abbrv-doi-hyperref}

\bibliography{Nagano_ISMAR2024}

\appendix % You can use the `hideappendix` class option to skip everything after \appendix

\end{document}